\documentclass[twocol]{ametsoc}
\journal{jas}

\usepackage{moreverb}
\usepackage{graphicx,overpic,natbib}
\usepackage{bm,url,color,colortbl,xcolor,soul}
\usepackage{url}
\usepackage{multirow}
\usepackage{booktabs} 
\usepackage[toc,page]{appendix}
\usepackage{float}
\usepackage{wrapfig}
\usepackage{amsmath}
\newcommand\BibTeX{{\rmfamily B\kern-.05em \textsc{i\kern-.025em b}\kern-.08em
T\kern-.1667em\lower.7ex\hbox{E}\kern-.125emX}}
\newcommand{\Fig}[1]{Figure~\ref{#1}}
\newcommand{\Tab}[1]{Table~\ref{#1}}
\newcommand{\m}{\rm{m}}
\newcommand{\s}{\rm{s}}
\newcommand{\xx}{\bm{x}}
\newcommand{\EQ}{\begin{equation}}
\newcommand{\EN}{\end{equation}}
\newcommand{\Eq}[1]{Equation~(\ref{#1})}

\graphicspath{{./fig/}{./png/}}

\bibpunct{(}{)}{;}{a}{}{,}

\title{Condensational and collisional growth of cloud droplets in a turbulent environment\footnotemark[2]}
\authors{Xiang-Yu Li\correspondingauthor{Xiang-Yu Li, Department of Meteorology and Bolin Centre for Climate Research, Stockholm University, Stockholm, Sweden}}
\affiliation{Department of Meteorology and Bolin Centre for Climate Research, Stockholm University, Stockholm, Sweden;\\
Nordita, KTH Royal Institute of Technology and Stockholm University, 10691 Stockholm, Sweden;\\
Swedish e-Science Research Centre, www.e-science.se, Stockholm, Sweden;\\
JILA and Laboratory for Atmospheric and Space Physics, University of Colorado, Boulder, CO 80303, USA}
\email{xiang.yu.li@su.se,~ $ $Revision: 1.413 $ $}
\extraauthor{Axel Brandenburg}
\extraaffil{
Nordita, KTH Royal Institute of Technology and Stockholm University, 10691 Stockholm, Sweden;\\
JILA and Laboratory for Atmospheric and Space Physics, University of Colorado, Boulder, CO 80303, USA;\\
Department of Astronomy, Stockholm University, SE-10691 Stockholm, Sweden}
\extraauthor{Gunilla Svensson}
\extraaffil{Department of Meteorology and Bolin Centre for Climate Research, Stockholm University, Stockholm, Sweden;\\
Swedish e-Science Research Centre, www.e-science.se, Stockholm, Sweden}
\extraauthor{Nils E.\ L.\ Haugen}
\extraaffil{SINTEF Energy Research, 7465 Trondheim, Norway; \\
Department of Energy and Process Engineering, NTNU, 7491 Trondheim, Norway}
\extraauthor{Bernhard Mehlig}
\extraaffil{Department of Physics, Gothenburg University, 41296 Gothenburg, Sweden}
\extraauthor{Igor Rogachevskii}
\extraaffil{Department of Mechanical Engineering, Ben-Gurion Univ. of the Negev, P. O. Box 653, Beer-Sheva 84105, Israel;\\
Nordita, KTH Royal Institute of Technology and Stockholm University, 10691 Stockholm, Sweden}

\abstract{
We investigate the effect of turbulence on the combined condensational and collisional growth
of cloud droplets by means of high resolution direct numerical
simulations of turbulence and a superparticle approximation for droplet dynamics and collisions.
The droplets are subject to  turbulence as well as gravity,
and their collision and coalescence efficiencies are taken to be unity.
We solve the thermodynamic equations governing temperature, water-vapor mixing ratio,
and the resulting supersaturation fields together
with the Navier-Stokes equation.
We find that the droplet-size distribution broadens with
increasing Reynolds number and/or mean energy dissipation rate.
Turbulence affects the condensational
growth directly through supersaturation fluctuations, and it
influences collisional growth indirectly through condensation.
Our simulations show for the first time that, in the absence of the mean updraft cooling,
supersaturation fluctuation-induced broadening of droplet-size distributions
enhances the collisional growth. This is contrary to classical (non-turbulent) condensational
growth, which leads to a growing mean droplet size, but a narrower droplet-size distribution.
Our findings, instead, show that condensational growth facilitates
collisional growth by broadening the size distribution in the tails at
an early stage of rain formation. With increasing
Reynolds numbers, evaporation becomes stronger. This
counteracts the broadening effect due to condensation at
late stages of rain formation.
Our conclusions are consistent with results of laboratory
experiments and field observations, and show that supersaturation
fluctuations are important for precipitation.
}

\begin{document}
\maketitle

\clearpage

\section{Introduction}

It has been suggested that warm rain accounts for about $30\%$ of the total
amount of rain and for $70\%$ of the total rain area in the tropics, which
plays an important
role in regulating the vertical water and energy transport of the tropical
atmosphere \citep{lau2003warm}. Its rapid formation has puzzled
the cloud microphysics community for about 70 years.
The observed timescale of warm rain formation is known to be about 20 minutes
\citep{stephens2007near}, which is much shorter than the theoretically predicted
timescale of 8 hours \citep{1955_Saffman} and 60 minutes in simulations of classical
adiabatic parcel models \citep{jonas1996turbulence}.
Condensational and collisional growth
determine the formation of warm rain. In the absence of turbulence,
condensational growth is effective for cloud condensation nuclei and cloud
droplets smaller than 15 $\mu$m in radius. Since the growth rate is
inversely proportional to the radius, condensational growth leads
to a narrow width of the droplet-size distribution.
The gravity-generated collisional growth in isolation becomes important only when
the mean radius of droplets is larger than $\sim40\,\mu$m,
and the collision efficiency becomes large enough
for collisional growth.
Thus, there is a size gap
of $15\,\mu$m--$40\,\mu$m where neither condensation nor collision--coalescence
drives the growth \citep{pruppacher2012microphysics, 2011_lamb, Grabowski_2013}.
Therefore, the effect of turbulence on condensational and collisional growth
has been proposed to overcome this size gap \citep{1955_Saffman, shaw_2003, KPE07,
DBB12, Grabowski_2013}.
In the meteorology community, the process of collision--coalescence
is also referred to as collection \citep{berry_1974}, while in the astrophysical community,
this process is referred to as coagulation \citep{li2017effect, johansen2017forming}. Since
we assume unit collision and coalescence efficiency, we use the terminology
collision in the present study.

\cite{1955_Saffman} showed that turbulent mixing enhances the droplet-collision
rate, following an idea of \cite{Smo17}.
They found
that this rate is proportional to the mean energy dissipation rate of
the turbulent flow. The calculation assumes
that the droplets are so small (about $10\,\mu$m in radius)
that inertial effects \citep[see][for a review]{GM16} are negligible.
More recently it has become clear that inertial effects
can significantly increase the collision rate for larger droplets, with larger Stokes numbers
\citep{Sun97,Fal02,CC05,Wil06,SA08,bec2010intermittency, gustavsson2011distribution,gustavsson2014relative,Gus14e,meibohm2017relative}.
These predictions are in good agreement with direct numerical simulations (DNS)
of droplet dynamics in turbulence \citep{Bhatnagar2018b,Bhatnagar2018a}, but the effect applies only to droplets
that are large enough that they can frequently detach from the flow,
due to the formation of caustics \citep{Wil05}. This requires Stokes numbers of order unity.

\citet{reuter1988collection}, \citet{grover1985effect},
\citet{pinsky2004collisions}, and \citet{pinsky2007collisions,
pinsky2008collisions} also suggested that turbulence may cause
a substantial enhancement of the collision rate, yet \citet{koziol1996effect}
found that turbulence only has a moderate effect on the collision rate.
This may partially be due to small Stokes numbers.

Recently it has become feasible to study the
condensational and collisional growth using DNS.
Most DNS studies of droplet collisions in turbulence
\citep{franklin2005collision, ayala_2008, rosa2013kinematic, chen2016cloud, woittiez2009combined}
record collision frequencies but do not allow the droplets to coalesce and grow. It is then
not possible to study how the droplet-size distribution develops.
Nevertheless, those works revealed that turbulence enhances the collision rate, and the effect is
larger for larger mean energy dissipation rates. The value of the Reynolds number, by contrast,
was found to be of  secondary importance.

\citet{franklin2008warm}, \citet{xue2008growth}, and \citet{Wang_2009}
investigated the collision--coalescence
processes by solving the Smoluchowski equation together with the Navier-Stokes
equation using DNS. They found that the size distribution of cloud droplets
is significantly enhanced by turbulence.
\citet{2016Onishi} extended the collision-rate model of \citet{Wang_2009} and
performed DNS at higher Reynolds number, where a Reynolds
number dependency was obtained.
Using a Lagrangian collision-detection method, \cite{chen2018turbulence}
found that turbulence strongly affects the broadening of the size distribution.
\cite{li2017effect} showed that, in the absence of condensation, turbulence enhances
the collision--coalescence process. They also found that this enhancement effect
is sensitive to the initial width of the droplet-size distribution.

The effect of turbulence on condensational growth has been explored
intensively. Since turbulence affects the temperature field and
spatial distribution of the water-vapor mixing ratio,
the supersaturation field determined
by temperature and water mixing ratio is inevitably affected by
turbulence.
\cite{srivastava1989growth} criticized the use of volume-averaged supersaturation
and proposed adopting the local supersaturation field
to calculate the condensational growth of cloud droplets.
This is a prototype of supersaturation fluctuations.
To investigate how local supersaturation fluctuations affect
the condensational droplet growth in the cloud core, \cite{vaillancourt2002microscopic} solved
the thermodynamical equations that govern the supersaturation using DNS
in the presence of a turbulent flow, taking into account the mean updraft
cooling, gravitational settling, droplet inertia, and
latent-heat release.
\cite{vaillancourt2002microscopic} concluded that
the width of the droplet-size distribution decreases
as the turbulent mean energy dissipation rate increases
and attributed this to the decrease in the decorrelation
time of the supersaturation fluctuation.
\cite{2009_Lanotte}, \cite{2015_Sardina}, and
\cite{siewert2017statistical} performed DNS
for condensational growth using a slightly simpler model
that accounts for supersaturation fluctuations
but not for details of the thermodynamics. They found that the size
distribution broadens as the Reynolds number increases.
\cite{paoli2009turbulent} found that the entrainment-induced supersaturation
fluctuations broaden the droplet-size distribution.
Their study is based on stochastically forced temperature and vapor fields.
\cite{grabowski2017broadening} and \cite{abade2018broadening} came to a similar conclusion
using a turbulent adiabatic-parcel model.
\citet{li2018cloud} confirmed that the droplet-size distribution
broadens with increasing Reynolds number and is insensitive to
the mean energy dissipation rate of turbulence.
Field observations of the supersaturation fluctuations and the droplet
size distribution \citep{Siebert17,yang2018cloud, desai2019search}
also support the idea that supersaturation
fluctuations due to turbulence lead to broadening of droplet-size
distribution.

Most of the previous DNS studies only considered either condensational
growth or collisional growth. The combined condensational and collisional
growth has rarely been investigated.
Recently, \cite{saito2017turbulence} studied the combined processes
using DNS. They found that the width of the droplet-size distribution increases
as the turbulence intensity increases. However, they did not discuss
whether it is the Reynolds number or the mean energy dissipation
rate that matters for the broadening.
\cite{chen2018bridging} employed the same model as \cite{saito2017turbulence}
and concluded that droplet-size distributions broaden with increasing
mean energy dissipation rate. However, they did not study the dependency
of the droplet-size distribution upon the Reynolds number.
Indeed, several works \citep{2009_Lanotte,2015_Sardina,siewert2017statistical,li2018cloud}
suggested that condensational growth is sensitive to the Reynolds number.
Collisional growth, however, is mainly affected by the mean energy dissipation rate
\citep{ayala_2008, chen2016cloud, li2017effect}.

In this paper, we investigate the effect of turbulence on the combined condensational
and collisional growth of cloud droplets at high Reynolds numbers using DNS of turbulence.
We strive to investigate whether/how supersaturation
fluctuations-induced broadening of droplet-size distributions
affect the collision--coalescence process, and thereby the
warm rain formation.
For the dynamics of the local temperature and the local water-vapor mixing ratio
we use the same model as \cite{vaillancourt2002microscopic}, \cite{saito2017turbulence},
and \cite{chen2018bridging}, excluding the mean updraft cooling (see below).
Details of our implementation are given in \cite{li2018cloud}.
The droplet dynamics in our simulations is coupled to the turbulence through Stokes force. The droplets
are also subject to gravitational settling.  DNS of the combined problem poses
formidable challenges.  Therefore we use a stochastic
Monte-Carlo approximation, the  superparticle method
\citep{Dullemond_2008, Shima09,li2017eulerian,Unterstrasser17},
for the collision--coalescence process. Strengths and weaknesses of the method were discussed by
\cite{li2018fluctuations}.
Since we focus on the impact of turbulence on droplet growth we omit the effect of cooling
due to a mean updraft.

We first investigate how condensational and collisional processes affect each other
through thermodynamics and droplet dynamics.
Second, we explore how the combined condensational and collisional droplet growth
depends on the mean energy dissipation rate and upon the Reynolds number.
We focus on the droplet-size distribution,
which is  the key to cloud-climate feedback and precipitation \citep{shaw_2003}.
We show that collisional growth is enhanced by the appearance of a broadening
tail of the droplet-size distribution through supersaturation fluctuations.

\section{Numerical model}

The equations governing the Eulerian fields and condensation
are the same as the standard ones \citep{vaillancourt2002microscopic},
and their implementation is described in \citet{li2018cloud}.
For the collision--coalescence dynamics we use the superparticle method,
which has been validated in \citet{li2017eulerian}.
The {\sc Pencil Code} is used for all the simulations.

\subsection{Eulerian fields and condensation}
\label{sec:model}

We use the standard equations for fluid density $\rho(\xx, t)$, fluid  velocity $\bm{u}(\xx, t)$, temperature $T(\xx, t)$,
and water-vapor mixing ratio $q_v(\xx, t)$:
\begin{equation}
\label{eq:continuity}
  {\partial\rho\over\partial t}+{\bm{\nabla}}\cdot(\rho\bm{u})=S_\rho ,
\end{equation}
\begin{equation}
{D\bm{u}\over D t}=\bm{F}
-\rho^{-1}{\bm{\nabla}} p
  +\rho^{-1} {\bm \nabla} \cdot (2 \nu \rho \mbox{\boldmath ${\sf S}$})+B\bm{e}_z+{\bm S}_u ,
\label{turb}
\end{equation}
\EQ
\label{eq:tt}
{D T\over Dt}
=\kappa\nabla^2 T+\frac{L}{c_{\rm p}}C_d ,
\EN
\EQ
{D q_v\over Dt}=D\nabla^2 q_v-C_d,
\label{mixingRatio}
\EN
where turbulence is driven by a stochastic forcing function
${\bm F}$ \citep[see][for details]{Haugen_etal_2004PhRvE},
$D/Dt=\partial/\partial t+\bm{u}\cdot{\bm{\nabla}}$ denotes the advective derivative,
and ${\sf S}_{ij}={\textstyle\frac{1}{2}}(\partial_j u_i+\partial_i u_j)
-{\textstyle{1\over3}}\delta_{ij}{\bm{\nabla}}\cdot\bm{u}$ is
the rate-of-strain tensor (subtracting the divergence makes it traceless), $p$ and $\rho$ are gas pressure and density,
$L$ is the latent heat. The parameters  $D$ and $\kappa$ are the diffusivities of
water vapor and  temperature.
The source terms  $S_\rho$ and $\bm{S}_u$ in Equations~(\ref{eq:continuity}) and (\ref{turb})
describe  mass transfer between the droplets and the humid air due to condensation and evaporation.
In our case, the mass transfer is small relative to the total air mass,
and the fraction of liquid to gaseous mass is also low.
Therefore we neglect these terms.
The pressure $p$ and the density $\rho$ are related to each other by
an adiabatic equation of state, $p=\rho c_{\rm s}^2/\gamma$,
where $c_{\m s}$ is the sound speed set in the code, $\gamma=c_{\rm p}/c_{\rm v}=7/5$, with
$c_{\rm p}=1005\,{\rm J}\,{\rm kg}^{-1}\,{\rm K}^{-1}$ being the
specific heat at constant pressure and $c_{\rm v}$ the
specific heat at constant volume.
For the kinematic viscosity and the thermal diffusivity of air, we use
$\nu=\kappa=1.5\times10^{-5}\m^2\s^{-1}$.
Furthermore, $D=2.55\times10^{-5}\m^2\s^{-1}$ is the water vapor diffusivity
and $L=2.5\times10^6\,{\rm J}\,{\rm kg}^{-1}$ is the latent heat.

The buoyancy force $B(\xx,t)$ is determined by the temperature $T(\xx,t)$
through $B=g(T^{\prime}/T+\alpha q_v^{\prime}-q_l)$,
where $g=9.81\m\s^{-2}$ is the gravitational acceleration,
$T^{\prime}=T-T_{\rm env}$ is the temperature fluctuation
with respect to the environmental temperature $T_{\rm env}=293\,$K,
$\alpha=0.608$ is the expansion coefficient,
$q_v^{\prime}=q_v-q_{v,\rm{env}}$ is the fluctuation of the water-vapor mixing ratio
\citep{2011_lamb,Kumar14}, with
$q_{v,\rm{env}}=0.01\,{\rm kg}\,{\rm kg}^{-1}$; see also \citet{li2018cloud}.
This follows the common approach \citep{vaillancourt2002microscopic} in
that it uses the Boussinesq approximation to describe the term $B e_z$
in Equations~(\ref{turb}),
assuming that density variations
are negligible except when multiplied by the gravitational acceleration;
see for example \cite{mehaddi2018inertial}.
This requires that temperature gradients are small.
Our implementation is slightly different from the classical Boussinesq
approximation, where $\bm{\nabla}\cdot\bm{u}=0$ is assumed.
Here, we use instead the full continuity equation (\ref{eq:continuity}).

Both $T$ and $q_v$ are affected by droplets via the condensation rate $C_d$
\citep{vaillancourt2001microscopic, li2018cloud},
\EQ
C_d\left(\xx,t\right)
= \frac{4\pi\rho_l G}{\rho_a}
\langle
s\left(\xx,t\right)r\left(t\right)\rangle \overline{n}\,.
\label{condensationRate}
\EN
The average $\langle \cdots\rangle$ represents a local
average over droplets at position $\xx$ and of volume $\eta^3$,
where $\eta$ is the Kolmogorov length, and
$\overline{n}=N_{\triangle}/(\Delta x)^3$ is the number of
droplets $N_{\triangle}$ per grid volume $(\Delta x)^3$.
The parameters are: liquid-water density $\rho_l=1000\,{\rm kg}\,{\rm m}^{-3}$,
reference mass density of dry air $\rho_a=1\,{\rm kg}\,{\rm m}^{-3}$,
condensation parameter $G=1.17\times10^{-10}\m^2\s^{-1}$,
supersaturation $s(\xx,t)=q_v/q_{vs}(T)-1$,
saturated
water-vapor mixing ratio $q_{vs}(T)=e_s(T)/R_v \rho_0 T$ with
gas constant $R_v=461.5\,{\rm J}\,{\rm kg}^{-1}\,{\rm K}^{-1}$.
Finally, $e_s$ is the saturation pressure obtained from the
Clausius-Clapeyron equation \citep{yau1996short,Gotzfried17},
$e_s(T)=c_1 \exp(-c_2/T)$. For the two constants we choose
$c_1=2.53\times10^{11}\,$Pa and $c_2=5420\,$K, as in \cite{li2018cloud}.

\subsection{Droplet dynamics and collisions: the superparticle algorithm}
We approximate the droplet dynamics using
the superparticle method \citep{Dullemond_2008, Shima09, Johansen_2012, li2017eulerian, li2017effect}.
In this approach, several identical microscopic droplets are represented by a superparticle.
Each superparticle is assumed a certain volume and is thus assigned
a droplet number density, $n_i$.
The position of superparticle $i$ is denoted by $\bm{x}_i$ and obeys
\begin{equation}
  \frac{d\bm{x}_i}{dt}=\bm{V}_i,
\end{equation}
where $\bm{V}_i$ is the velocity of the superparticle. The acceleration obeys Stokes law
\begin{equation}
  \frac{d\bm{V}_i}{dt}=\frac{1}{\tau_i}(\bm{u}-\bm{V}_i)+\bm{g},
\end{equation}
where $\tau_i$ is the Stokes time \citep{li2017effect},
$\bm{u}$ is the fluid velocity at $\bm{x}_i$, and $\bm{g}$
is the gravitational acceleration.
The value of $\tau_i$ of superparticle $i$ is given by
\begin{equation}
\label{response_time}
  \tau_i=2\rho_{\rm d} r_i^2/[9\rho\nu \, (1+0.15\,{\rm Re}_i^{2/3})].
\end{equation}
Here the term $1+0.15\,{\rm Re}_i^{2/3}$ \citep{Schiller33, Marchioli08}
is due to particle Reynolds numbers,
${\rm Re}_i=2r_i|\bm{u}-\bm{V}_i|/\nu$.
We adopt this term since the largest particle
Reynolds number becomes large,
when $r$ exceeds values of around $r=100\,\mu\rm{m}$,
and the linear Stokes drag does not hold.
Droplet collisions are modeled as follows
\citep{Shima09,Johansen_2012,li2017eulerian, Unterstrasser17}.
When two superparticles reside in the same grid cell, the probability of collision
between one droplet in a superparticle with a droplet in another
superparticle during time step $\Delta t$ is $p_{c}=\tau_{c}^{-1}\Delta t$.
The collision time $\tau_{c}$ is determined by
\begin{equation}
	\tau_{c}^{-1}=\sigma_{c} n_j\left|\bm{V}_{i}
	-\bm{V}_{j}\right| E_{c}.
\label{tauij1}
\end{equation}
Here $\sigma_{c}=\pi(r_i+r_j)^2$ is the
cross section between two colliding droplets.
The collision efficiency $E_{c}$ is treated as unity.
We refer to \citet{li2017eulerian} and \citet{li2017effect} for details of the algorithm.

The cloud system is very dilute, with a typical
mean number density of about $10^8\rm{m^{-3}}$ in stratocumulus clouds. Considering a
$1\,\rm{m}$ cubic domain in the cloud core, the number of droplets is $10^8$. The typical
Kolmogorov length scale is about $\eta=1\,\rm{mm}$. To resolve the Kolmogorov scale of
cloud-like turbulence in clouds, about $(1\,\rm{m}/1\,\rm{mm})^3=10^9$ grid points are
needed in DNS. This means that there is only 1 cloud droplet in a cube with volume $(10\eta)^3$,
i.e., 1 droplet in every 10 grid boxes in DNS. With such a dilute system, stochasticity
is argued to become important for the collision--coalescence process
\citep{Kos05, wilkinson16}. The inherent stochastic
property of the superparticle approach renders it an ideal method to study the collision-
coalescence process in cloud system \citep{Dziekan17,Unterstrasser17, grabowski2018modeling}.
This realization emerged as an important consensus
among Shima, Unterstrasser, Dziekan, and others during a recent
workshop\footnote{\url{http://ww2.ii.uj.edu.pl/~arabas/workshop_2019/}}
on ``Eulerian vs.\ Lagrangian methods for cloud microphysics,''
held in Krakow in April 2019.
Comparing with the direct Lagrangian collision--coalescence
detection method, the superparticle method is computationally
more efficient because it avoids having to
follow each droplet individually \citep{Shima09, Johansen_2012, li2017eulerian}.

\begin{table*}[t!]
\caption{Parameter values used in the different simulation runs.
Here ``cond'' refers to condensation; ``coll'' refers to collision;
``both'' refers to combined condensation and collision.
Runs~C1 and E1 are reference runs that agree with Runs~A and C of
\cite{li2018cloud} (condensation only) and Runs~C2 and E2 are similar
to Runs~A and C of \cite{li2017effect} (collisions only, except that here
the initial mean number density of droplets is $n_0=2.5\times10^8\,\rm{m}^{-3}$).
To allow for a comparison with the reference runs, we chose the
parameters for Runs A, B, C, D, and E to be the same as those by
\cite{li2018cloud}. These authors studied {\em only} condensation.
Here collisions and condensation are treated together.
The amplitude of the random forcing $F_0$,
the lateral size of the cubic simulation box $L_x$,
the number of grid cells $N_{\rm grid}$,
and the eddy turnover time $\tau_L$ were defined
in Section~2.c.
}
\centering
\setlength{\tabcolsep}{3pt}
\begin{tabular}{|l|c|c|c|c|c|c|c|c|c|c|r|}
\hline
Run &  $F_0$ & $L_x\,(\rm{m})$ & $N_{\rm grid}$ & $N_{\rm s}$& Processes & $u_{\rm rms}$ ($\rm{m} \, \rm{s}^{-1}$) & $\mbox{\rm Re}_{\lambda}$ & $\bar{\epsilon}$ ($\rm{m}^2\rm{s}^{-3}$)& $\eta$ ($\rm{mm}$) & $\tau_{\eta}$ ($\rm{s}$) & $\tau_L$ ($\rm{s}$) \\
\hline
A & $0.007$ & 0.200 & $128^3$ & $244140$& both &0.10 &44 & 0.005 & 0.92  & 0.056 & 0.67 \\ 
B & $0.014$ & 0.150 & $128^3$ & $244140$& both &0.14 &45  & 0.019 & 0.65  & 0.028 & 0.35 \\ 
C & $0.020$ & 0.125 & $128^3$ & $244140$& both &0.16 &45  & 0.039 &0.54  & 0.020 & 0.25 \\ 
C1 & $0.020$ & 0.125 & $128^3$ & $244140$& cond    &0.16 &45  & 0.039 &0.54  & 0.020 & 0.25 \\ 
C2 & $0.020$ & 0.125 & $128^3$ & $244140$& coll   &0.16 &45  & 0.039 &0.54  & 0.020 & 0.25 \\ 
D & $0.020$ & 0.250 & $256^3$ & $1953120$& both&0.22 &78  & 0.039  & 0.54  & 0.020 & 0.37 \\ 
E & $0.020$ & 0.500 & $512^3$ & $15624960$& both&0.28 &130  & 0.039  & 0.54  & 0.020 & 0.58 \\ 
E1 & $0.020$ & 0.500 & $512^3$ & $15624960$& cond &0.28 &130  & 0.039  & 0.54  & 0.020 & 0.58 \\ 
E2 & $0.020$ & 0.500 & $512^3$ & $15624960$& coll &0.28 &130  & 0.039  & 0.54  & 0.020 & 0.58 \\ 
\hline
\multicolumn{12}{p{0.9\textwidth}}{}
\end{tabular}
\label{Swarm_Rey}
\end{table*}

The superparticle approach has been validated against the Smoluchowski
equation in both pure gravity cases \citep{Shima09, Unterstrasser17, li2017eulerian}
and in turbulent cases \citep{li2017eulerian}. Good agreement was observed.
Its stochasticity was investigated by \citet{Dziekan17}, who found that
the superparticle approach reproduces stochastic coalescence when $N_{\rm p}/N_{\rm s}\leq9$,
where $N_{\rm p}$ is the total number of physical particles and $N_{\rm s}$
is the total number of superparticles.
This suggests that the superparticle
approach does indeed capture the stochasticity of the Lagrangian
collision--coalescence detection method sufficiently accurately when
$N_{\rm p}/N_{\rm s}$ becomes sufficiently small.
Comparison with the direct Lagrangian collision--coalescence detection method
is still under investigation in the cloud microphysics community.
Nevertheless, \citet{onishi2015lagrangian}
compared the direct collision--coalescence detection method with the Smoluchowski
equation and found excellent agreement. This suggests that the
superparticle approach converges to the direct Lagrangian collision--coalescence
detection method.

\subsection{DNS}
The present study builds upon our earlier simulations of
condensational growth \citep{li2018cloud} and collisional growth \citep{li2017effect}.
Here we treat both processes {\em together} in order to determine how the two
mechanisms interact. Our numerical setup is the
same as in \cite{li2018cloud}, apart from the fact that we now include
collisional growth. Details of our DNS solver are given in \cite{li2017eulerian,li2018cloud}.
To mimic the diluteness of the cloud system, $N_{\rm s}/N_{\rm grid}=0.1$
is adopted, which is within the convergence range $N_{\rm s}/N_{\rm grid}\le0.05$
\citep{li2017effect}.
This also ensures that the tails of $f(r,t)$ are statistically
converged for larger values of $N_{\rm s}$, and thus larger ${\rm Re}_\lambda$.
More importantly, we keep $N_{\rm p}/N_{\rm s}=2$ so that
the stochasticity of the superparticle approach is correctly represented,
which is well within the limit $N_{\rm p}/N_{\rm s}\le9$
determined by \citet{Dziekan17}.
Log-normal initial distributions with different width
($\sigma=0$, 0.02, 0.05, and 0.1) are employed in the
present study.

To investigate how the time evolution of droplet-size distribution depends
on the Taylor micro-scale Reynolds number ${\rm Re}_\lambda$
($\equiv u_{\rm rms}^2 \sqrt{5/(3\nu\bar{\epsilon})}$) and
the mean energy dissipation rate $\bar{\epsilon}$, we performed
high resolution simulations with different domain sizes $L_x$
and different non-dimensional forcing amplitude $F_0$, which is a
prefactor in each Fourier component of wavevector $\bm{k}$ given by
$F_0 c_{\rm s}\,(|\bm{k}|c_{\rm s}/\Delta t)^{1/2}$.
We choose $\bm{k}$ from a narrow band of wavevectors such that
$|\bm{k}|L_x/2\pi\approx3$.

Our results are summarized in Table \ref{Swarm_Rey}.
To elucidate the combined effect of condensational and collisional growth,
we use our earlier simulations as references; see \cite{li2018cloud}
for condensational growth and \cite{li2017effect} for collisional growth.
The corresponding runs are also listed in the Table.

We run simulations for 10 minutes even for the largest Reynolds number
(Runs E, E1, and E2 with ${\rm Re}_\lambda=130$). 1066600 time steps
with $dt=3.405\times10^{-5}\,\rm{s}$ are integrated in a wall-clock time of
$24\times11$ hours using 4096 CPUs. This corresponds to 1034 eddy turnover times.
The droplet Stokes time is about $1.5\times10^{-3}\,\rm{s}$ for the
smallest droplet. Therefore, the time is well-resolved.

\begin{figure*}[t!]\begin{center}
\includegraphics[width=\textwidth]{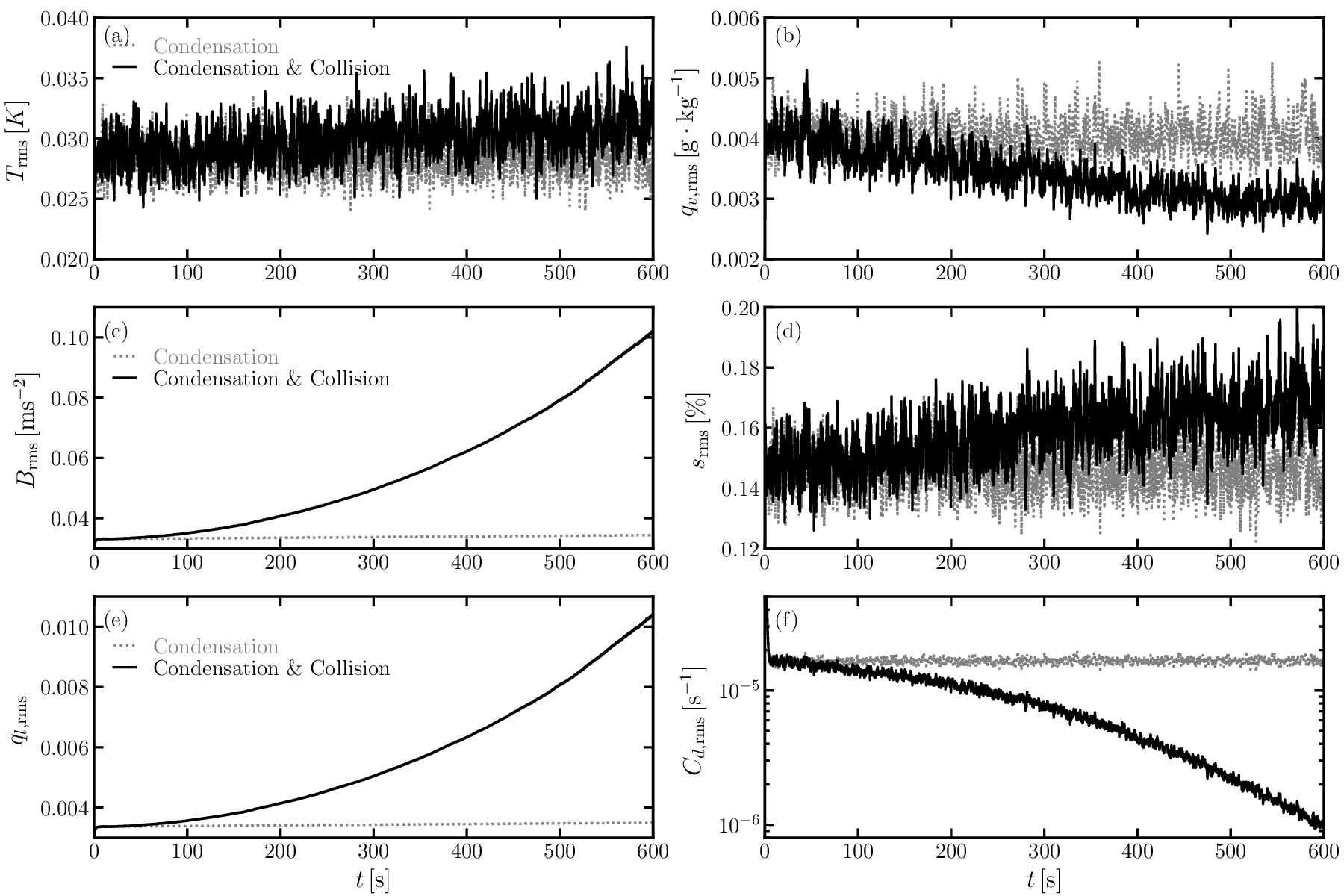}
\end{center}
\caption{
Comparison of rms values of various thermodynamic quantities in the presence (absence)
of collisions shown as solid (dotted) lines, corresponding to Run~C (C1).
Condensation is included in both cases.
(a) $T_{\rm rms}(t)$; (b) $q_{v},\rm{rms}(t)$; (c) $B_{\rm rms}(t)$; (d) $s_{\rm rms}(t)$;
(e) $q_{l,\rm rms}(t)$; and (f) $C_{d,\rm rms}(t)$.
}
\label{timeSeries_gravRe50_cond_coa}
\end{figure*}

\subsection{Diagnostics}

The tail of droplet-size distribution determines
warm rain formation and precipitation.
We characterize the length of the tail of $f(r,t)$
by the normalized moments of $r$ as \citep{li2017eulerian}
\begin{equation}
  a_\zeta=\left({M_\zeta}/{M_0}\right)^{1/\zeta},
\label{azeta}
\end{equation}
where $M_\zeta=\int_{0}^\infty f\,r^\zeta \,{\rm d}r$
is the $\zeta$th moment of $r$.
The case of $\zeta\to\infty$ corresponds to the maximum of $r$
and the case $\zeta=1$ corresponds to the mean radius, $\bar{r}$.
In practice, we choose $\zeta=24$ as a reasonably stable compromise
to quantify the end of the tail of the distribution.

The relative dispersion of $f(r,t)$ is characterized by
$\sigma_r/\bar{r}$, where $\sigma_r$ is the standard deviation
of the droplet size and $\bar{r}$ is the mean radius \citep{igel2017importance}.
The standard deviation of $f(r,t)$ is given by
\EQ
\sigma_r=\sqrt{a_2^2-a_1^2}
\EN
in terms of the normalized moments of $r$, defined in \Eq{azeta}.
Thus $\sigma_r/\bar{r}=(a_2^2-a_1^2)^{1/2}/a_1$.

\section{Results}

\subsection{Comparison between cases with condensational growth, collisional growth, and with both}
\label{ComparisonDifferentCases}

\begin{figure}[t!]\begin{center}
\includegraphics[width=0.5\textwidth]{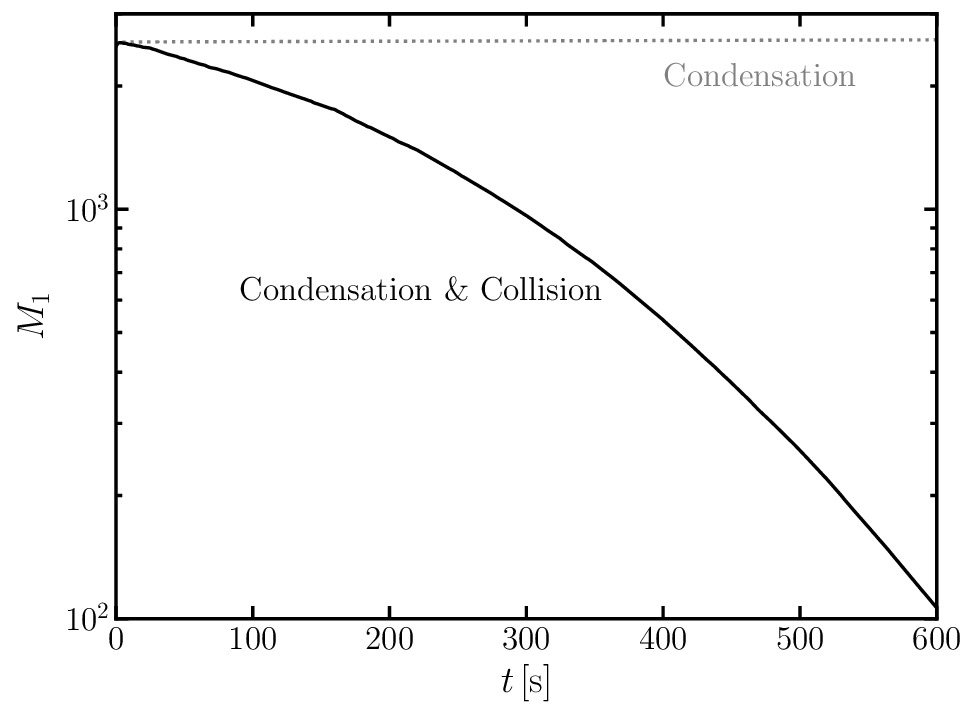}
\end{center}
\caption{
Evolution of $M_1$ for simulations shown in
\Fig{timeSeries_gravRe50_cond_coa}.
}
\label{a1_gravRe50_cond_coa}
\end{figure}

Condensational growth of cloud droplets is affected by supersaturation fluctuations
\citep{2009_Lanotte, 2015_Sardina, siewert2017statistical,
grabowski2017broadening, li2018cloud, abade2018broadening}.
These fluctuations are governed
by temperature $T(\xx,t)$ and by the water-vapor mixing ratio $q_v(\xx,t)$.
We first investigate how collision impacts these quantities and therefore the
condensational growth.
\Fig{timeSeries_gravRe50_cond_coa} shows the time series of $T_{\rm rms}(t)$,
$q_{v,\rm rms}(t)$, $B_{\rm rms}(t)$, $s_{\rm rms}(t)$, $q_{l,\rm rms}(t)$,
and $C_{d,\rm rms}(t)$ with and without collisions.
We see that the collision--coalescence process affects
the fluctuations of these quantities to different degrees.
Both $T_{\rm rms}(t)$ and $s_{\rm rms}(t)$ increase
due to the collision--coalescence process while
$q_{v,\rm rms}(t)$ decreases slightly at the late
stage of rain formation.
This can be explained by the response of $C_{d,\rm rms}(t)$
to the collision--coalescence process.
After about $t=100\,\rm{s}$,
the collision--coalescence process becomes dominant.
Since $s_{\rm rms}(t)$ only increases slightly, $C_{d,\rm rms}(t)$
is determined by $M_1$, as shown in \Eq{condensationRate}.
\Fig{a1_gravRe50_cond_coa} shows that $M_1$ decays
rapidly as collision--coalescence becomes important.
This results in a decease of $C_{d,\rm rms}(t)$
by about an order of magnitude.
The decreasing $C_{d,\rm rms}(t)$ leads to a positive
feedback on $T_{\rm rms}(t)$ and $s_{\rm rms}(t)$,
and a negative feedback on $q_{v,\rm rms}(t)$.
The buoyancy force $B$ is determined by temperature fluctuations $T^{\prime}$, water-vapor
mixing ratio fluctuations $q^{\prime}_v$, and the liquid-water mixing ratio $q_l$.
The collision--coalescence process
leads to more intense local variations of $q_l$, which result in larger values of $q_{l,\rm{rms}}$.
Therefore, $B_{\rm rms}(t)$ is enhanced by the collision--coalescence process through $q_l$.
In our simulations, however, the enhanced $B$ does not affect the
flow field since the random forcing overwhelms the buoyancy
force in our simulations.
Thus, collisional growth does not impact the condensational growth in
the present DNS.
This may change when the volume stirring of the flow is replaced
by buoyant driving. Such driving could be more realistic, especially on larger
scales that cannot be accessed in current state-of-the-art DNS.

\begin{figure*}[t!]\begin{center}
\includegraphics[width=\textwidth]{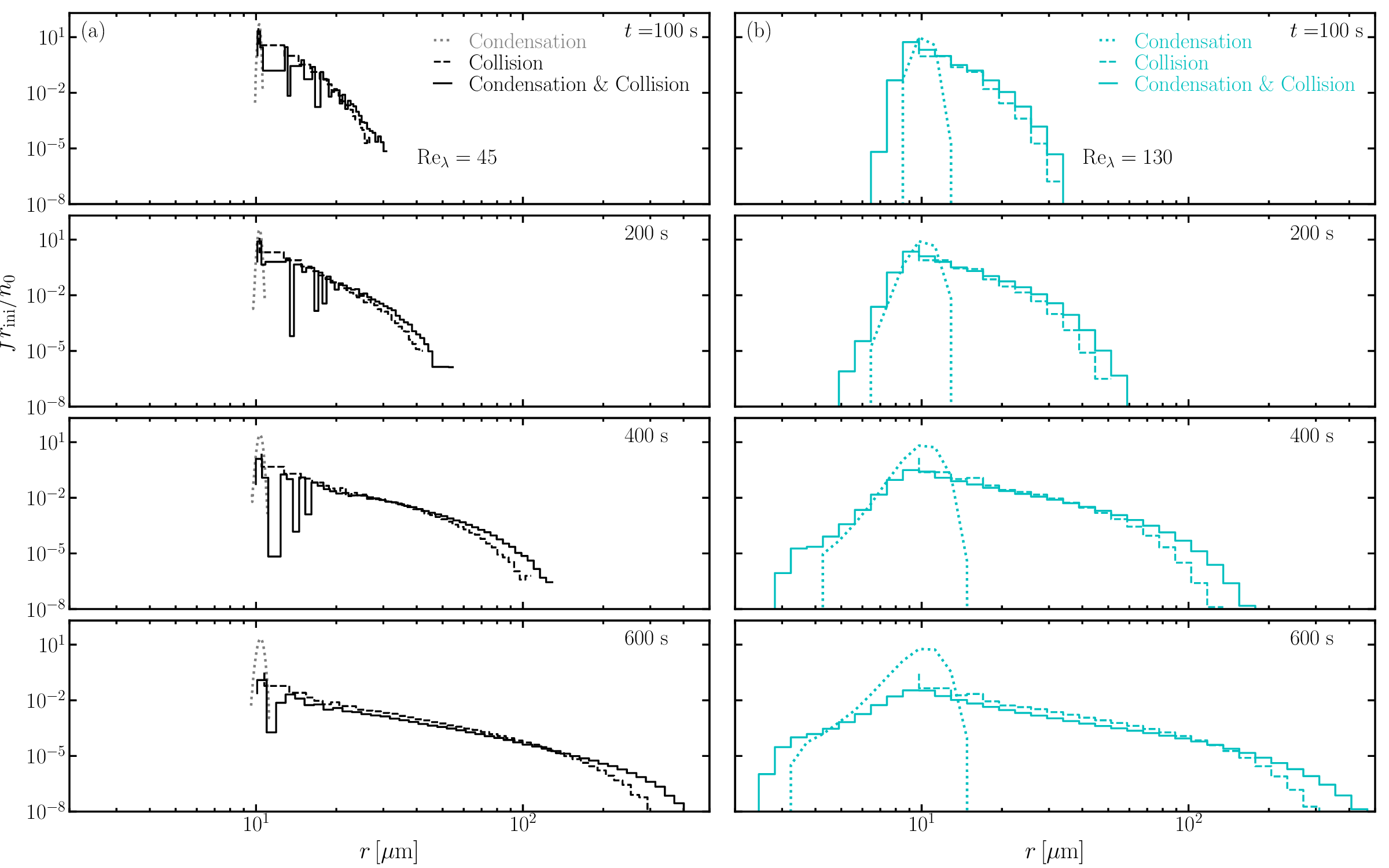}
\end{center}\caption{
Comparison of droplet-size distributions for three cases:
condensational growth [Runs~C1 and E1 in Table~\ref{Swarm_Rey}; \cite{li2018cloud}], dotted lines;
collisional growth [Runs~C2 and E2 in Table~\ref{Swarm_Rey}; \cite{li2017effect}], dashed lines, and
the combined processes (Runs~C and E in Table~\ref{Swarm_Rey}), solid lines.
(a) $\rm{Re}_\lambda=45$; (b) $\rm{Re}_\lambda=130$.
}
\label{f_gravRe50_cond_coa_comp}
\end{figure*}

Next, we investigate how condensational growth affects the collisional growth
by comparing the time evolution of the droplet-size distribution for three
different cases: condensation only, collision only, and the combined process.
\Fig{f_gravRe50_cond_coa_comp}(a) shows the comparison
of droplet-size distributions when ${\rm Re}_\lambda=45$
and $\bar{\epsilon}=0.039\,\rm{m}^2 \rm{s}^{-3}$.
For the case with only condensation, the width of the droplet-size distribution
increases from a monodispersed initial distribution.
When comparing the tail of the size distribution
between the cases of collision only and that of the combined process,
we see that the broadening from the condensational
growth facilitates the collisional growth.
The combined condensational and collisional growth leads
to large tails of the size distribution.
In \Fig{f_gravRe50_cond_coa_comp}(b),
we show the corresponding result for $\rm{Re}_\lambda=130$.
At $t=600\,\rm{s}=10\,\rm{min}$,
the radius of the droplet reaches about $400\,\mu\rm{m}$,
which is almost the size of falling raindrops. This timescale is
close to the observed timescale for warm rain formation.
It is worth noting that for the combined process,
the droplet-size distribution exhibits an obvious transition from
condensational growth to collisional growth, as shown by
the dip in the droplet-size distribution.
We recall that the radius of all droplets is initially $r_{\rm ini}\equiv10\,\mu\rm{m}$.
After the first collision, the droplet grows to twice the mass, giving
a radius of $12.6\,\mu{\rm m}$.
Condensational growth leads
to a few large droplets close from the initially monodispersed
$10\,\mu\rm{m}$ droplet distribution, which triggers the collision--coalescence process.
For the case of ${\rm Re}_\lambda=130$ (cyan curves), the dips are less abrupt.
This is due to the fact that larger
value of ${\rm Re}_\lambda$ lead to stronger supersaturation
fluctuations, which thus generate more large droplets.

\begin{figure}[t!]\begin{center}
\includegraphics[width=0.5\textwidth]{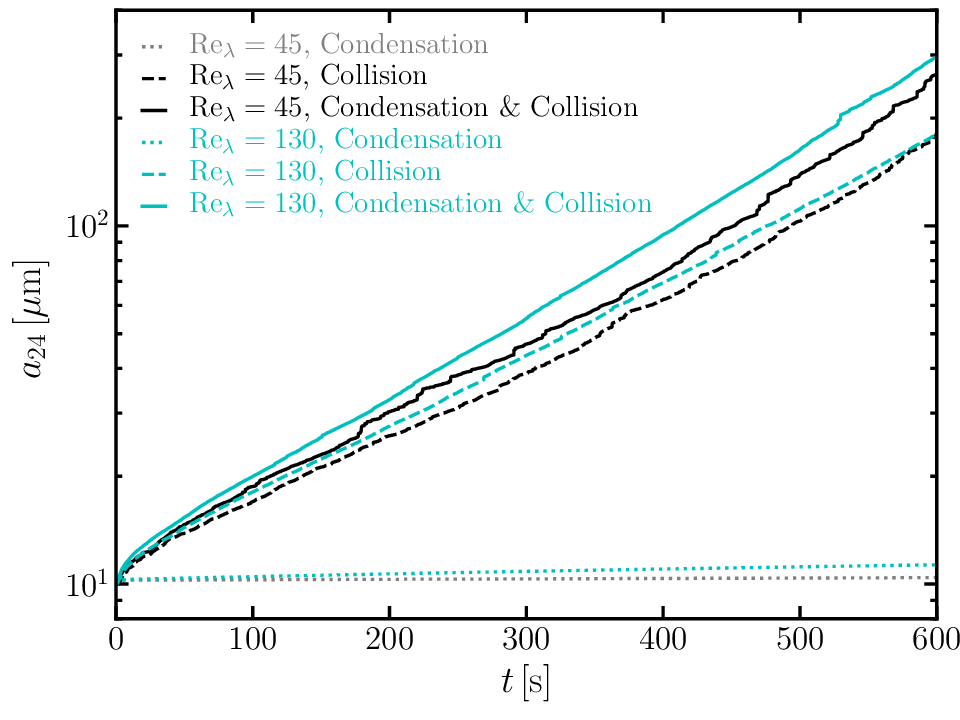}
\end{center}\caption{
Evolution of $a_{24}$ for different configurations:
condensation only (dotted lines); collision--coalescence
only (dashed lines); condensation and collision--coalescence
(solid lines). Black curves are for the cases with $\rm{Re}_\lambda
=45$ and the cyan ones for $\rm{Re}_\lambda=130$.
}
\label{a24_gravRe50_cond_coa_comp}
\end{figure}

To see how the tail of $f(r,t)$ depends on
${\rm Re}_\lambda$ for different configurations,
we examine $a_\zeta$.
As shown in \Fig{a24_gravRe50_cond_coa_comp},
$a_{24}$ is insensitive to ${\rm Re}_\lambda$ when condensation
is excluded, which is consistent with previous
studies \citep{chen2018turbulence, li2017effect}.
However, when condensation is included, $a_{24}$ increases with increasing
${\rm Re}_\lambda$.
This demonstrates that the value of $\rm{Re}_\lambda$
affects collisional growth indirectly through condensation.
For cases with only condensation, $a_{24}$ is larger
for larger $\rm{Re}_\lambda$.

We have also investigated how different initial distributions affect
the combined condensational and collisional growth. It is found that
the condensation process makes the combined processes almost independent
of the initial distribution; see the appendix for details.

The collisional growth of cloud droplets is very sensitive to the
tails of droplet-size distributions.
A few large droplets can undergo a runaway collision--coalescence
process by collecting small droplets. The cumulative collision
time of these few large droplets is much shorter than the mean
collision time \citep{Kos05}. Thus, fluctuations play an important
role in collisional growth. Condensational growth due to supersaturation
fluctuations facilitates
this runaway collision--coalescence process by generating the few large droplets
as demonstrated in this study.

\subsection{Effect of turbulence on combined condensational and collisional growth}

To study the effect of turbulence on the combined condensational and
collisional growth, we explore how the time evolution of the droplet-size distributions depends
on $\bar{\epsilon}$ and ${\rm Re}_\lambda$ in the case
when the growth of droplets is driven by both condensation
and collision--coalescence.
Several previous works \citep{2009_Lanotte,2015_Sardina,
siewert2017statistical, li2018cloud} showed that condensational growth
is enhanced with increasing ${\rm Re}_\lambda$, but
is insensitive to $\bar{\epsilon}$.
Collisional growth, however,
depends on $\bar{\epsilon}$ and is insensitive to ${\rm Re}_\lambda$
\citep{ayala_2008, chen2016cloud, li2017effect}.
Therefore, we expect that the combined condensational and collisional growth
depends on both ${\rm Re}_\lambda$ and $\bar{\epsilon}$.

\begin{figure*}[t!]\begin{center}
\includegraphics[width=\textwidth]{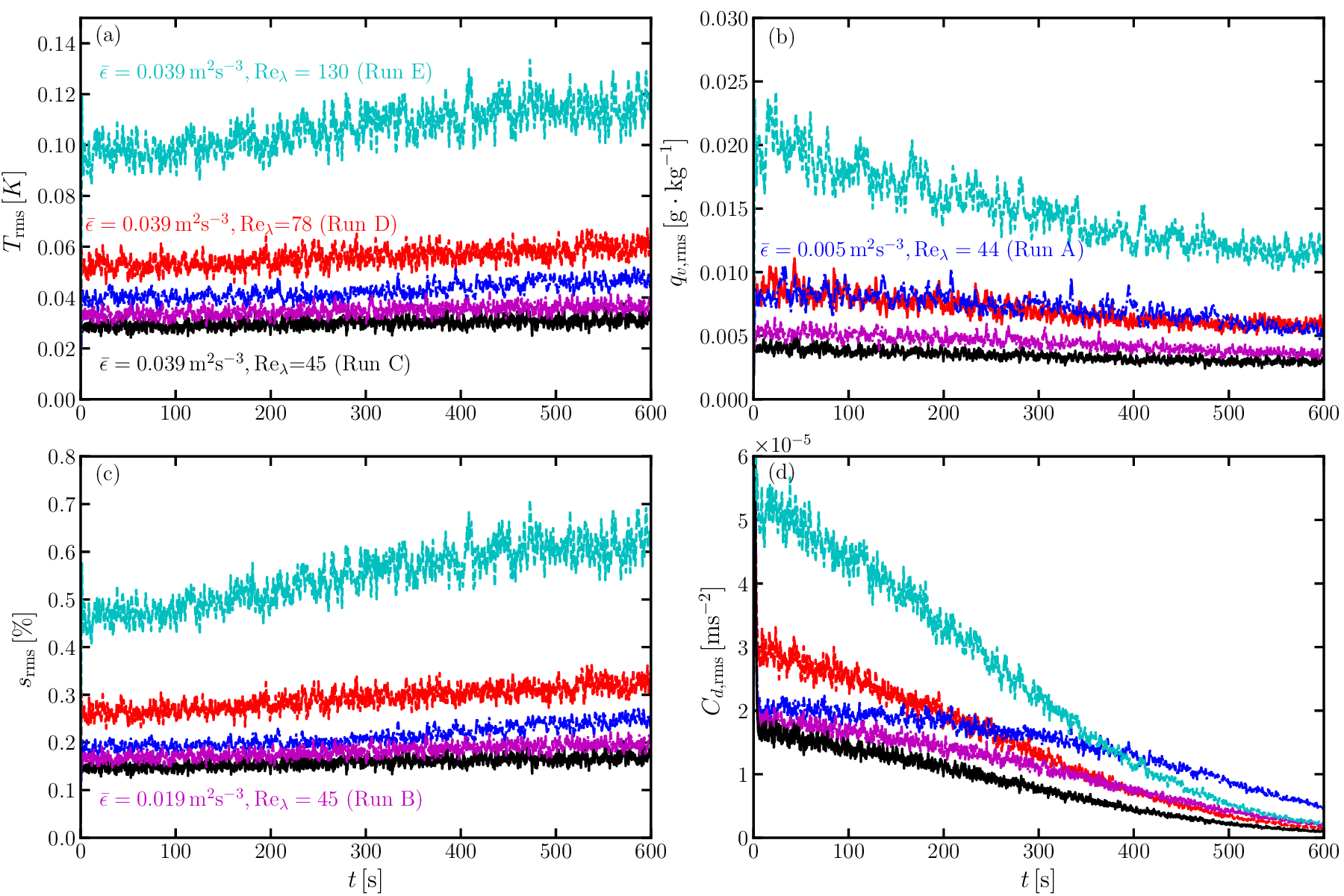}
\end{center}
\caption{
Evolution of the rms values of temperature (a), water-vapor mixing ratio (b),
supersaturation (c), and condensation rate (d)
for Runs~A (blue), B (magenta), C (black), D (red), and E (cyan).}
\label{timeSeries_cond_coa_comp}
\end{figure*}

Let us first inspect how the evolution of $T$, $q_v$, $C_d$, and $s$
depends on $\rm{Re}_\lambda$ and $\bar{\epsilon}$.
\Fig{timeSeries_cond_coa_comp} shows that the rms values of these
quantities increase as  $\rm{Re}_\lambda$ increases, but
they only depend weakly upon $\bar{\epsilon}$ (compare the blue,
magenta, and black lines for Runs~A, B, and C, respectively).
This result is consistent with the conclusion of
\cite{2015_Sardina} and \cite{li2018cloud},
where only the condensation/evaporation process was investigated,
but now it is also verified for the combined condensational and
collisional growth.

\begin{figure*}[t!]\begin{center}
\includegraphics[width=\textwidth]{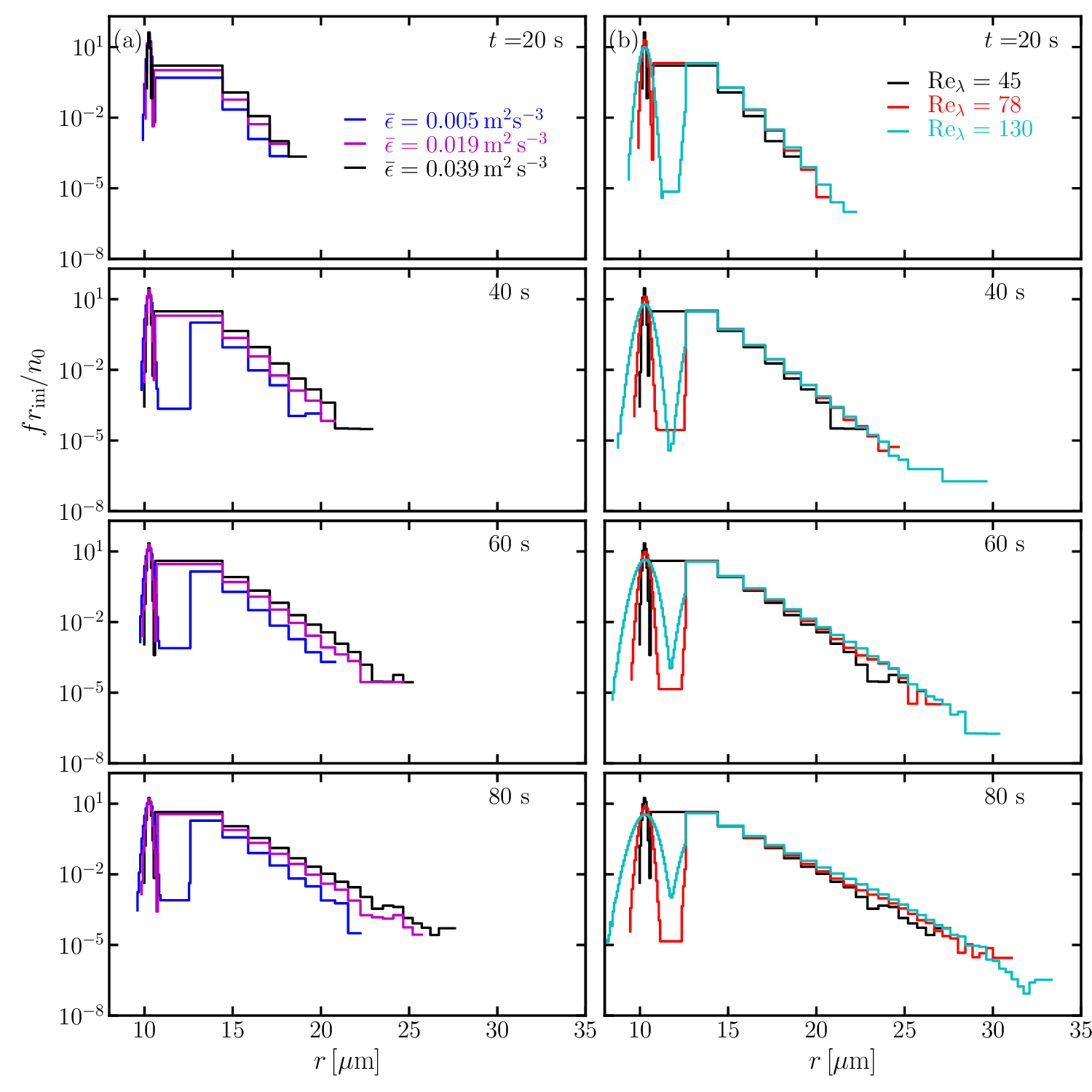}
\end{center}\caption{Droplet-size distributions for
(a) different $\bar{\epsilon}$ = $0.005\,\rm{m}^2\rm{s}^{-3}$ (blue solid lines),
0.019 (magenta solid lines), and 0.039
(black solid line) at fixed $\mbox{\rm Re}_{\lambda}=45$
(see Runs A, B, and C in Table~\ref{Swarm_Rey} for details)
and for (b) different $\mbox{\rm Re}_{\lambda}$ = 45 (black solid lines),
78 (red solid lines), and 130 (cyan solid line) at fixed
$\bar{\epsilon}$ = $0.039\,\rm{m}^2\rm{s}^{-3}$
(see Runs C, D, and E in Table~\ref{Swarm_Rey} for details).}
\label{f_cond_coa_comp_80s}
\end{figure*}

\Fig{f_cond_coa_comp_80s}(a) shows the time evolution of
the corresponding droplet-size distributions at an early stage
of rain formation.
Due to turbulence-induced supersaturation fluctuations,
the width of $f(r,t)$ broadens to a certain value.
The first peak at $r=10\mu{\rm m}$ and its width are almost the same for different
$\bar{\epsilon}$ at different times. The distributions exhibit the same
characteristics as those of the simulations without collisions in
Runs~C1 and E1 shown in \Fig{f_gravRe50_cond_coa_comp}.
We attribute this feature to the condensational growth and
its weak dependency on $\bar{\epsilon}$
\citep{2015_Sardina,li2018cloud}.
The tail of the droplet-size distribution becomes wider with increasing
$\bar{\epsilon}$, which is attributed to the dependency of collisional growth
on $\bar{\epsilon}$.
\Fig{f_cond_coa_comp_80s}(b), on the other hand, shows the time evolution of
the droplet-size distributions for different ${\rm Re}_\lambda$
at fixed $\bar{\epsilon}$. The first peak exhibits the same
shape and dependency on ${\rm Re}_\lambda$ as those in
\Fig{f_gravRe50_cond_coa_comp} where collisions were not included.
The distributions of small droplets become wider with increasing
${\rm Re}_\lambda$, which is due to the fact that
both evaporation and condensation are
enhanced with increasing ${\rm Re}_\lambda$.
This also indicates the strong spatial
inhomogeneity of the supersaturation field.
The tail of the droplet-size distribution broadens with increasing
${\rm Re}_\lambda$.
This is attributed to the condensational growth
and its induced collisions since collisional growth mainly depends
on $\bar{\epsilon}$.

\begin{figure*}[t!]\begin{center}
\includegraphics[width=\textwidth]{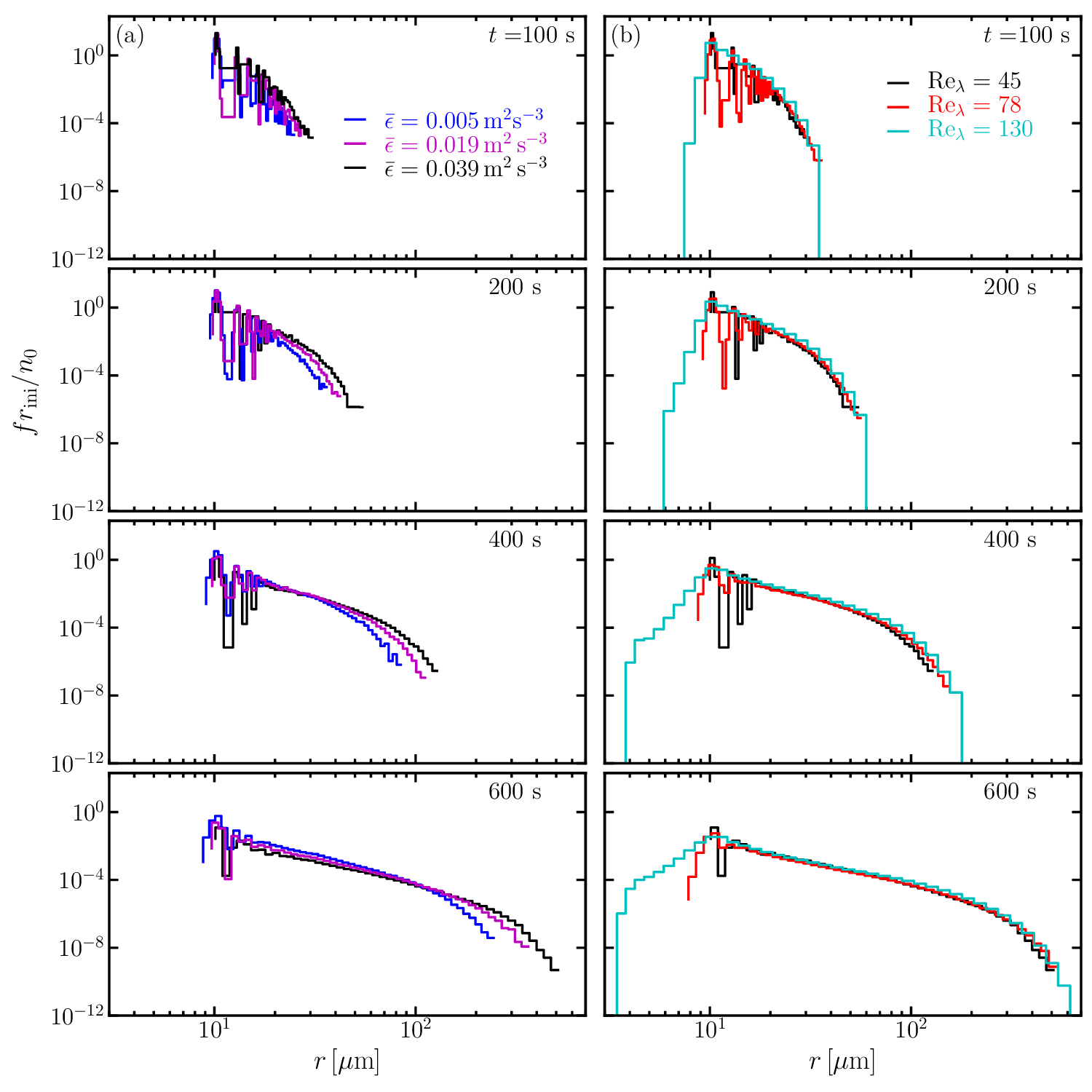}
\end{center}\caption{Same as \Fig{f_cond_coa_comp_80s} but
at late times.}
\label{f_cond_coa_comp}
\end{figure*}

\begin{figure}[t!]\begin{center}
\includegraphics[width=0.5\textwidth]{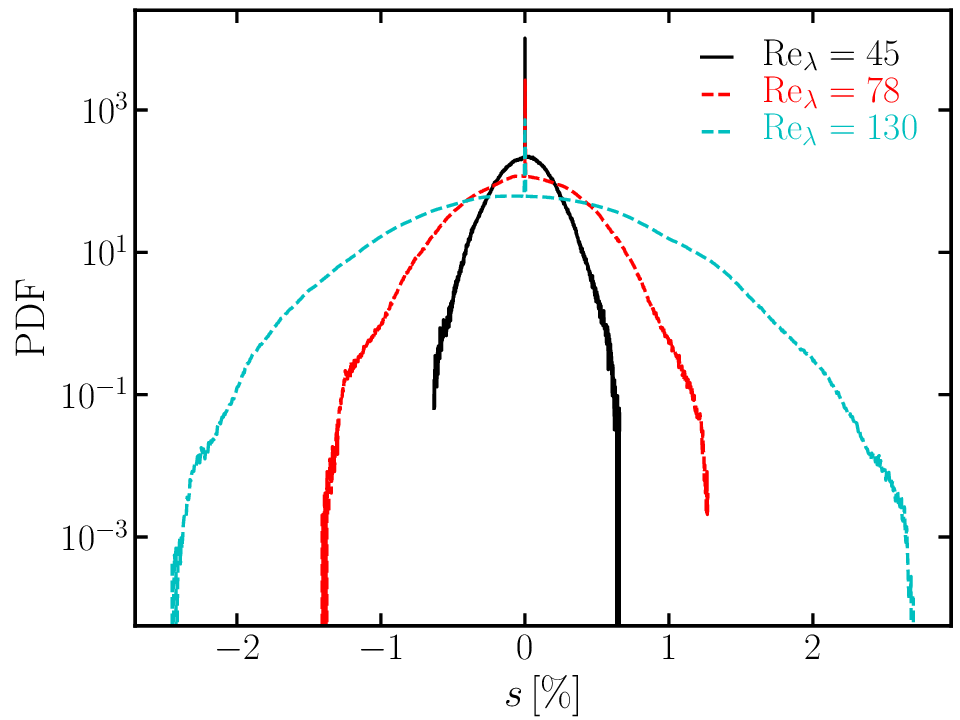}
\end{center}\caption{PDF of $s$ for different ${\rm Re}_\lambda$
at $t=600\,\rm{s}$.
Same simulations as in \Fig{f_cond_coa_comp}(b).}
\label{ssat_pdf_comp}
\end{figure}

\begin{figure*}[t!]\begin{center}
\includegraphics[width=\textwidth]{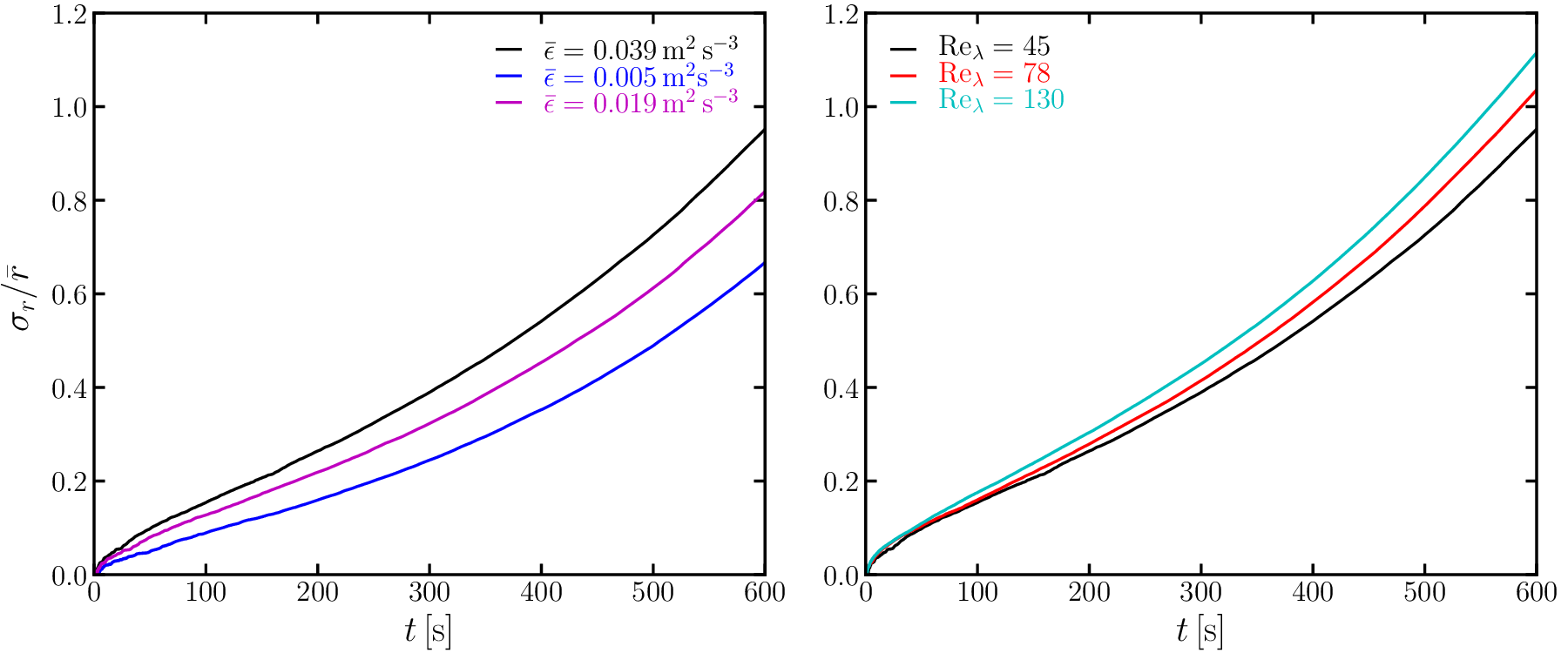}
\end{center}\caption{
Evolution of $\sigma_r/\bar{r}$ for the simulations shown in
\Fig{f_cond_coa_comp}.
}
\label{sigma_moments_cond_coll_0}
\end{figure*}

\begin{figure*}[t!]\begin{center}
\includegraphics[width=\textwidth]{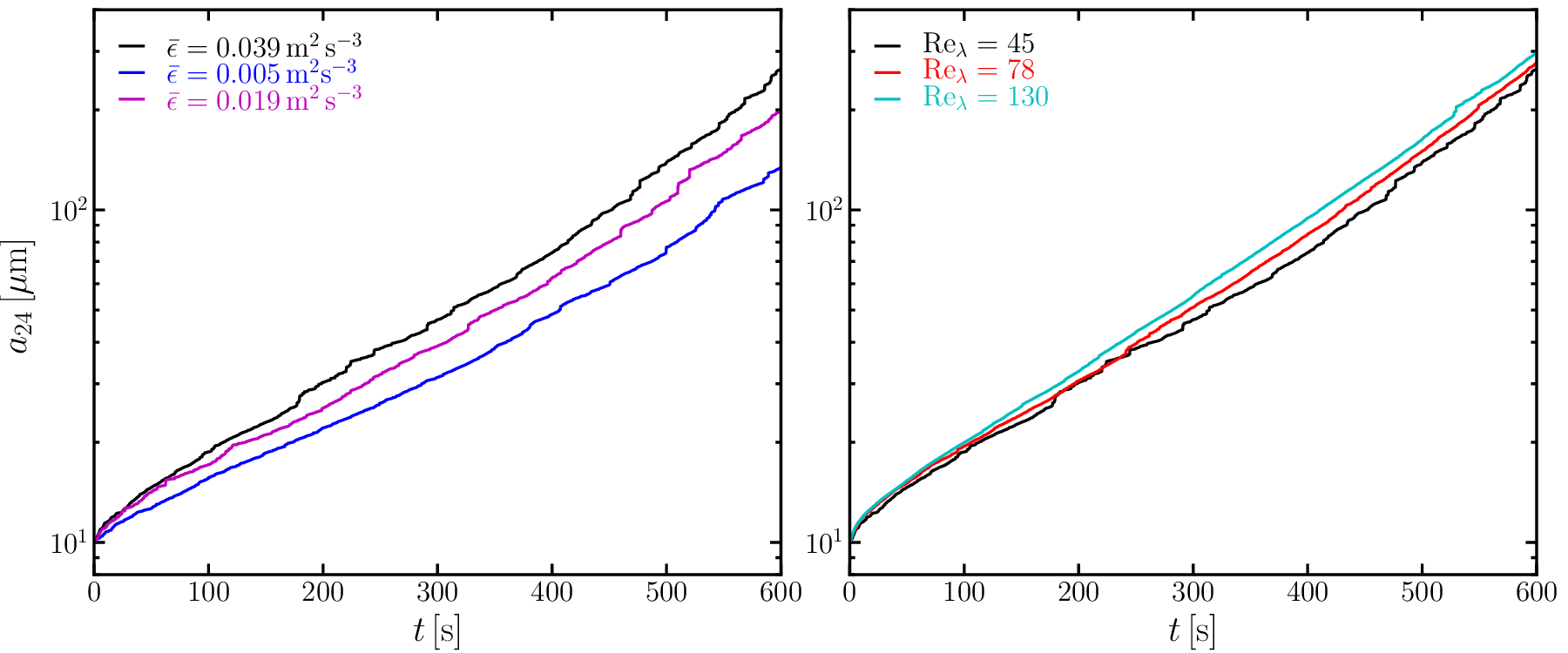}
\end{center}\caption{
Evolution of $a_{24}$ for the simulations shown in \Fig{f_cond_coa_comp}.
}
\label{a24_cond_coa_comp}
\end{figure*}

When the simulations ran for $600\,$s, we observe that
the ${\rm Re}_\lambda$-dependency becomes even stronger, as shown
in \Fig{f_cond_coa_comp}.
This is due to the fact that
evaporation results in smaller droplets, as can be seen
from the left tail of $f(r,t)$ in \Fig{f_cond_coa_comp}(b).
Larger values of ${\rm Re}_\lambda$ lead to stronger evaporation, and therefore
the broadening effect due to condensation at the early stage of rain formation
is counteracted by evaporation at the late stage.
The probability density function (PDF) of $s$ broadens significantly
with increasing values of ${\rm Re}_\lambda$.
This implies that there is stronger evaporation (negative $s$)
when ${\rm Re}_\lambda$ is larger; see \Fig{ssat_pdf_comp}.
From 60 to $80\,$s, the right tail
of $f(r,t)$ due to condensation does not broaden. Instead, its left tail now extends further.
The evolution of the dispersion of $f(r,t)$ is shown in
\Fig{sigma_moments_cond_coll_0}, where we observe
enhancement of $\sigma_r/\bar{r}$ with $\bar{\epsilon}$
and $\rm{Re}_\lambda$.
To characterize the tail of $f(r,t)$, we again inspect
the higher moments of $f(r,t)$.
As shown in \Fig{a24_cond_coa_comp}, the time evolution
of $a_{24}$ increases both with increasing $\bar{\epsilon}$
(due to collision--coalescence) and with increasing ${\rm Re}_\lambda$
(due to condensation). Within the parameter ranges of $\bar{\epsilon}$
and ${\rm Re}_\lambda$ in the present DNS,
the $\bar{\epsilon}$-dependency is more pronounced.
We noticed that there is exactly one particle per superparticles
for the smallest $f(r,t=600)$ for all the simulations. This excludes
the possibility that the wider tail of $f(r,t)$ is due to
a larger number of $N_{\rm s}$ for this case with the largest
${\rm Re}_\lambda$. This is consistent with our statement in section 2.c that
$N_{\rm p}/N_{\rm grid}=0.1$ is adopted in all simulations
to make sure that the tails of $f(r,t)$ are statistically converged.

\section{Discussion}

We have investigated how condensation and collision--coalescence processes
affect each other by comparing droplet-size distributions for three
cases: pure condensation, pure collision--coalescence, and the combined processes.
We found that condensational growth broadens the droplet-size distributions
in the initial phase of droplet growth, after which collisional growth
is triggered. The condensation-triggered collision is most pronounced for our largest
Reynolds number, $\rm{Re}_\lambda=130$.
In the present study, the collision--coalescence process
only enhances the buoyancy force and affects the
temperature, water-vapor mixing ratio, and supersaturation slightly.
Therefore, it does not influence the condensation process in the parameter
range explored.

We have also studied the combined condensational and collisional growth
at different $\bar{\epsilon}$ and $\rm{Re}_\lambda$. We observed that
the droplet-size distribution broadens both with increasing
$\bar{\epsilon}$ and with increasing $\rm{Re}_\lambda$.
The dependency on $\rm{Re}_\lambda$ can be explained as follows.
Several previous DNS studies \citep{ayala_2008, chen2016cloud, li2017effect}
showed that collisional growth
depends on $\bar{\epsilon}$ and is insensitive to $\rm{Re}_\lambda$.
The condensational growth, instead, strongly depends on $\rm{Re}_\lambda$ and is
insensitive to $\bar{\epsilon}$ \citep{li2018cloud}.
Also, in the present study, the comparison among cases
of pure condensation, pure collision, and the combined process demonstrates that
condensational growth triggers the collisional growth.
Therefore, we conclude that the $\rm{Re}_\lambda$ dependency
is caused by the condensation process,
which indirectly enhances the collisional growth.
The combined processes are also observed to depend on
$\bar{\epsilon}$, which is attributed to the dependency
of the collisional growth on $\bar{\epsilon}$. However,
the largest local $\bar{\epsilon}$ in warm clouds is about
$\bar{\epsilon} = 10^{-1}\,\rm{m}^2\rm{s}^{-3}$ \citep{siebert2013fine},
which is much smaller than the values
achieved in the laboratory and engineering applications \citep{shaw_2003}.
Its effect on collisional growth should be treated with caution.
The largest $\rm{Re}_\lambda$ is $130$ and the lateral size
of the domain is $L_x=0.5\,\rm{m}$ in the present DNS.
In reality, we have $\rm{Re}_\lambda\approx10^4$
\citep{siebert2006observations} in a cloud system
with a typical turbulence integral length scale of $100\,\rm{m}$,
which is two orders of magnitude larger than
the $\rm{Re}_\lambda$ in the present DNS.
It is expected that a higher $\rm{Re}_\lambda$ would lead to
larger supersaturation fluctuations \citep{grabowski2017broadening},
and therefore fast broadening of the size distribution,
which facilitates the collisional growth.
Our findings also support results of the laboratory experiment
of \cite{Chandrakar16} that supersaturation fluctuations
are likely of leading importance for precipitation
formation. Furthermore, we demonstrated numerically that
supersaturation fluctuations enhance the collisional
growth.

The classical treatment of condensational growth without turbulence,
and with constant supersaturation results in a larger mean
radius, but a narrower width of the size distribution.
This reduces the relative velocity of potentially
colliding pairs as they settle through the cloud. This implies {\em  slower} collisional growth.
Contrary to the classical treatment of condensational growth, our findings demonstrate
that the supersaturation fluctuation-induced condensational growth facilitates
the collisional growth by broadening the width of the droplet-size distribution.

\cite{chen2018bridging} compared droplet-size distributions for different
$\bar{\epsilon}$ when both condensation and collision were included.
They attributed the condensation-induced collision to the fact that
``\textit{condensational growth narrows the droplet size distribution (DSD) and provides
a great number of similar-sized droplets}'' \citep{chen2018bridging},
which is inconsistent with our finding that condensational growth
produces wider distributions with increasing $\rm{Re}_\lambda$ and therefore
facilitates the collisional growth.
However, we emphasize that there are two crucial differences compared
to our present model.
First, the mean updraft cooling included by \citet{chen2018bridging}
may suppress the supersaturation fluctuation-induced broadening of
the droplet-size distributions,
first found by \cite{sardina2018broadening}.
Mean updraft cooling may result in more similar-sized droplets generated by
the condensation process, which leads to their enhanced
collision rate.
Second, they included
hydrodynamic interactions between droplets.
This may modify the way how turbulence affects the collisional growth
discussed here.
In the present study, these two differences result in an
overestimation of the combined collisional and condensational growth.
When comparing the tail of $f(r,400)$
in \Fig{f_cond_coa_comp} with Fig.~1 of \citet{chen2018bridging},
our value is about $20\%$ larger.

In the present study, supersaturation fluctuations are caused purely by
the local condensation rate $C_d$ being affected by turbulence.
This results in extreme
supersaturation values, especially for the case of
the largest $\rm{Re}_\lambda$, as shown by the tail of the PDF
of $s$ in \Fig{ssat_pdf_comp}.
No equilibrium state of the supersaturation field is obtained;
see \Fig{timeSeries_cond_coa_comp}(c).
This is due to the following reason.
In the case of pure condensation,
supersaturation fluctuations become stationary
as they
relax to an equilibrium state \citep{2015_Sardina,li2018cloud}.
In the present study, the collision--coalescence process, however, leads
to a continuous growth of droplet sizes (no droplet breakup is included),
and alters the local concentration of droplets.
This prevents supersaturation fluctuations reaching an equilibrium state.

The continuous evolution of droplet-size distributions is sustained
by supersaturation fluctuations-induced broadening and the collision--coalescence
process.
In the pure condensation case, due to supersaturation fluctuations,
the standard deviation of the droplet surface area is proportional to
${\rm Re}_\lambda t^{1/2}$ \citep{2015_Sardina,li2018cloud}.
This leads to continuous broadening of droplet-size distributions
in the absence of mean updraft cooling \citep{sardina2018broadening}.
More importantly, the droplet-size distribution is further broadened
by the collision--coalescence process.
As discussed in section~2.a, stochastic
forcing is adopted in the present study, which
cannot sufficiently capture large scales of
turbulence. This is limited by the state-of-the-art
supercomputer power. This is why all the DNS studies
of the turbulence and cloud microphysics communities
\citep[e.g.][]{saito2017turbulence,chen2018bridging}
have employed volume-stirring.

Our study lends some support to the notion of ``lucky'' droplets \citep{Kos05},
first proposed by \citet{Telford1955}.
The lucky-droplet model assumes that there is a larger
droplet amongst many small ones to initiate the runaway
growth \citep{Kos05, wilkinson16}.
The question is where the first few lucky droplets come from.
\citet{Kos05} proposed that the first few lucky droplets
could be the result of giant condensation nuclei.
The present study indicates that
the first few lucky droplets
could result from condensational growth driven by supersaturation fluctuations
caused by turbulence.

\section{Conclusions}

We have found that
the growth of cloud droplets in warm clouds
is substantially affected by both the Reynolds number
and the mean energy dissipation rate.
The condensational growth is driven by supersaturation fluctuations. Supersaturation
fluctuations are governed by fluctuations of temperature and the water-vapor
mixing ratio, which were found to increase with increasing Reynolds number
\citep{2009_Lanotte,2015_Sardina,siewert2017statistical,li2018cloud}.
This results in a broadening of droplet-size distributions, which
is contrary to the classical understanding of condensational growth
that leads to a narrowing size distribution.
When the droplet-size distribution has reached a certain width, collisional growth
starts to dominate. It is then affected by
the mean energy-dissipation rate. In other words,
the value of the Reynolds number influences the collisional
growth {\em indirectly} through condensation.
Therefore, the combined condensational and collisional growth is
influenced by both the Reynolds number and the mean energy dissipation rate.
With the limited Reynolds numbers and the relatively
small domain size employed in the present DNS study,
we observed how the broadening of droplet-size distributions driven
by supersaturation fluctuations facilitates the collisional growth
at an early stage of rain formation. Evaporation becomes stronger
with increasing Reynolds number, which counteracts
the broadening of the droplet-size distribution with increasing
Reynolds number.

In the present study, the collision and coalescence efficiencies were
assumed to be unity, which may substantially
overestimate the collisional growth.
For example, the largest particle Reynolds number
is in excess of $500$ in some of our DNS, resulting
in droplet rebound or breakup, which can be
accounted for in the coalescence efficiency \citep[Page 406 of Chapter 9]{2011_lamb}.
This suggests the existence of an upper bound for
the enhancement of turbulence on collisional growth.
Since the turbulence-induced collision efficiency
is a very challenging problem \citep{Grabowski_2013}, it may be useful to incorporate a
robust scheme of collision efficiency in the superparticle approach.
Entrainment is also omitted, which
is supposed to cause strong supersaturation fluctuations.
Aerosol activation is not included in the present study.
Invoking all the cloud microphysical processes is computationally extremely
demanding -- even on modern supercomputers. We strive to achieve this in future studies.

Due to the aforementioned limitations, we have not attempted to compare
droplet-size distributions obtained from the current work
with observational data.
Such a step would make sense once we address the limitations
mentioned above and have a more realistic
representation of the large scales, where the flow is dominated by
convective driving instead of volume stirring, as in the present work.

\section*{Acknowledgements}

This work was supported through the FRINATEK grant 231444 under the
Research Council of Norway, SeRC,
the Israel Science Foundation governed by the
Israel Academy of Sciences (grant No.\ 1210/15),
the University of Colorado through its support of the
George Ellery Hale visiting faculty appointment,
the grant ``Bottlenecks for particle growth in turbulent aerosols''
from the Knut and Alice Wallenberg Foundation, Dnr.\ KAW 2014.0048,
and Vetenskapsr\aa{}det with grant number 2017-03865.
N.E.L.H.\ acknowledge the Research project Gaspro,
financed by the research council of Norway (267916).
The simulations were performed using resources provided by
the Swedish National Infrastructure for Computing (SNIC)
at the Royal Institute of Technology in Stockholm and
Chalmers Centre for Computational Science and Engineering (C3SE).
This work also benefited from computer resources made available through the
Norwegian NOTUR program, under award NN9405K.
The source code used for the simulations of this study, the {\sc Pencil Code},
is freely available on \url{https://github.com/pencil-code/}.

\appendix

\section{Independence of initial distribution}

\begin{figure}[t!]\begin{center}
\includegraphics[width=0.5\textwidth]{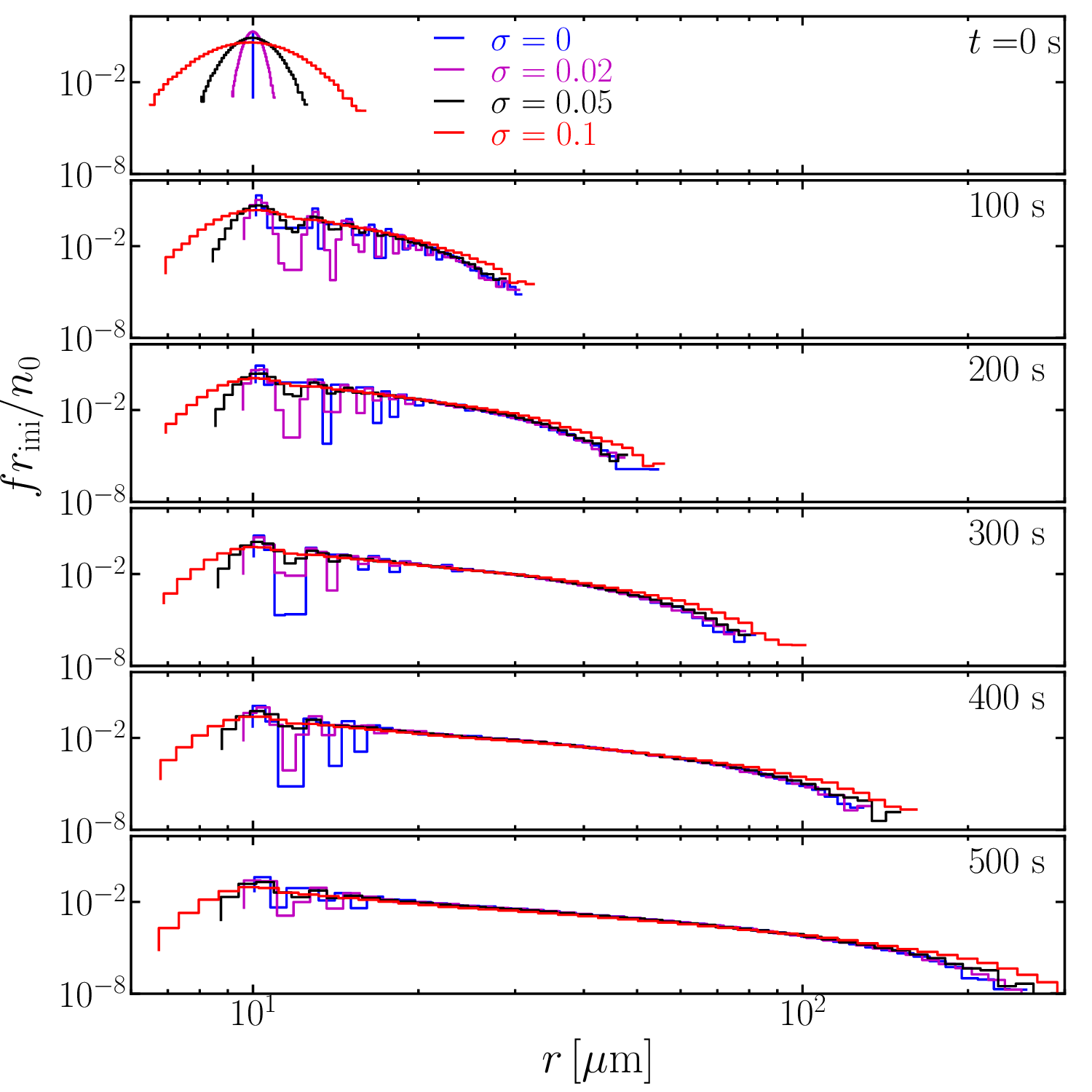}
\end{center}\caption{Comparison of droplet-size distributions
for different width $\sigma_{\rm ini}$;
see Runs C in \Tab{Swarm_Rey} for details of the numerical setup.
}
\label{f_cond_coa_comp_sigma}
\end{figure}

\begin{figure}[t!]\begin{center}
\includegraphics[width=0.5\textwidth]{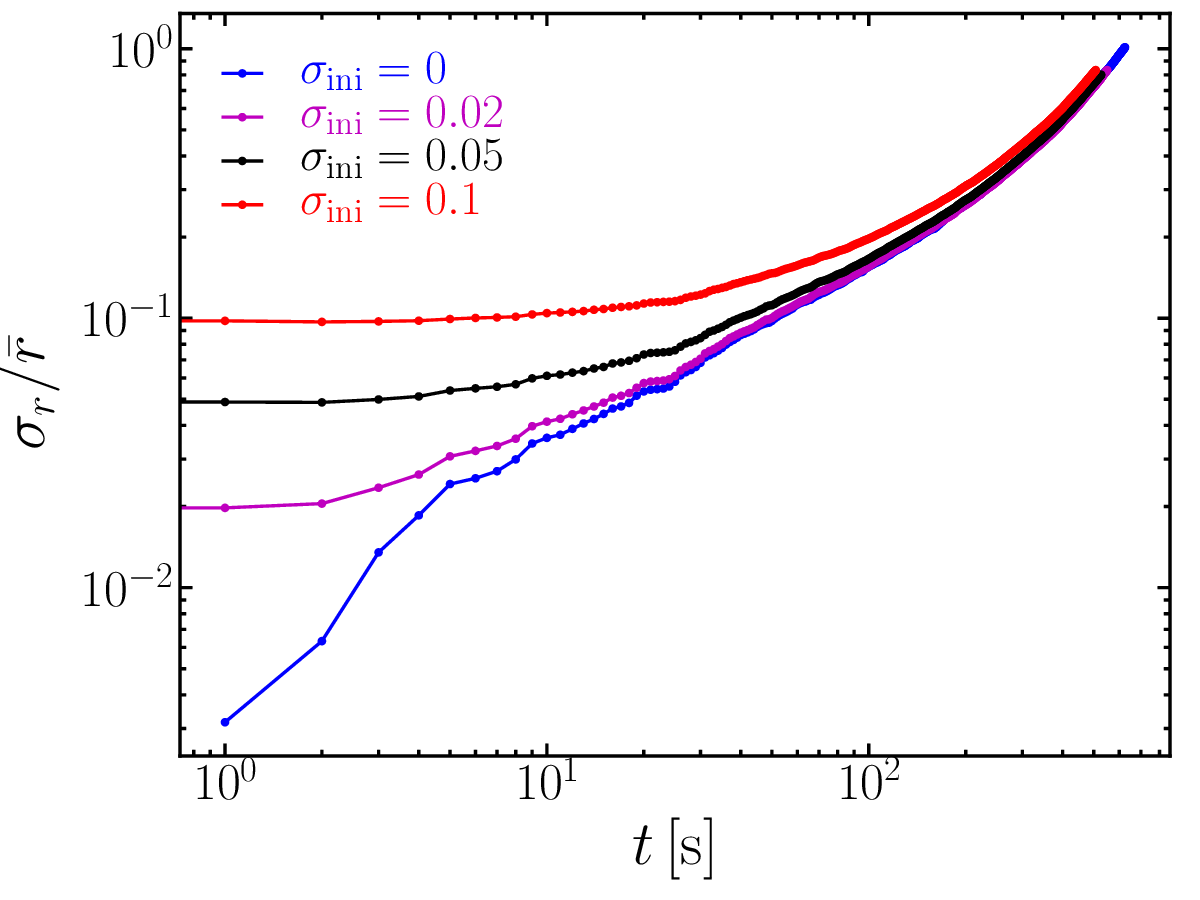}
\end{center}\caption{The corresponding relative dispersion
$\sigma_r/\bar{r}$ of droplet-size distributions
shown in \Fig{f_cond_coa_comp_sigma}.
}
\label{sigma_moments_cond_coll}
\end{figure}

In section~3a, we investigated how different initial distributions
affect the combined condensational and collisional growth.
In this appendix, we give further details and show how $f(r,t)$ depends on
$\sigma_{\rm ini}$.
\Fig{f_cond_coa_comp_sigma} shows that $f(r,t)$ is insensitive
to the width $\sigma_{\rm ini}$ of the initial size distribution.
As shown in \Fig{sigma_moments_cond_coll}, $\sigma_r/\bar{r}$
is insensitive to $\sigma_{\rm ini}$ at late times.
This is consistent with the behavior of $f(r,t)$ shown in
\Fig{f_cond_coa_comp_sigma}. Since $\sigma_r/\bar{r}$
only involves the second moment of the radius $r$, it is
not able to characterize the tail of $f(r,t)$. Therefore, we use
high moments of $r$ as defined in \Eq{azeta} of Section 3.
We show $\sigma_r/\bar{r}$ for comparison.

\begin{figure}[t!]\begin{center}
\includegraphics[width=0.5\textwidth]{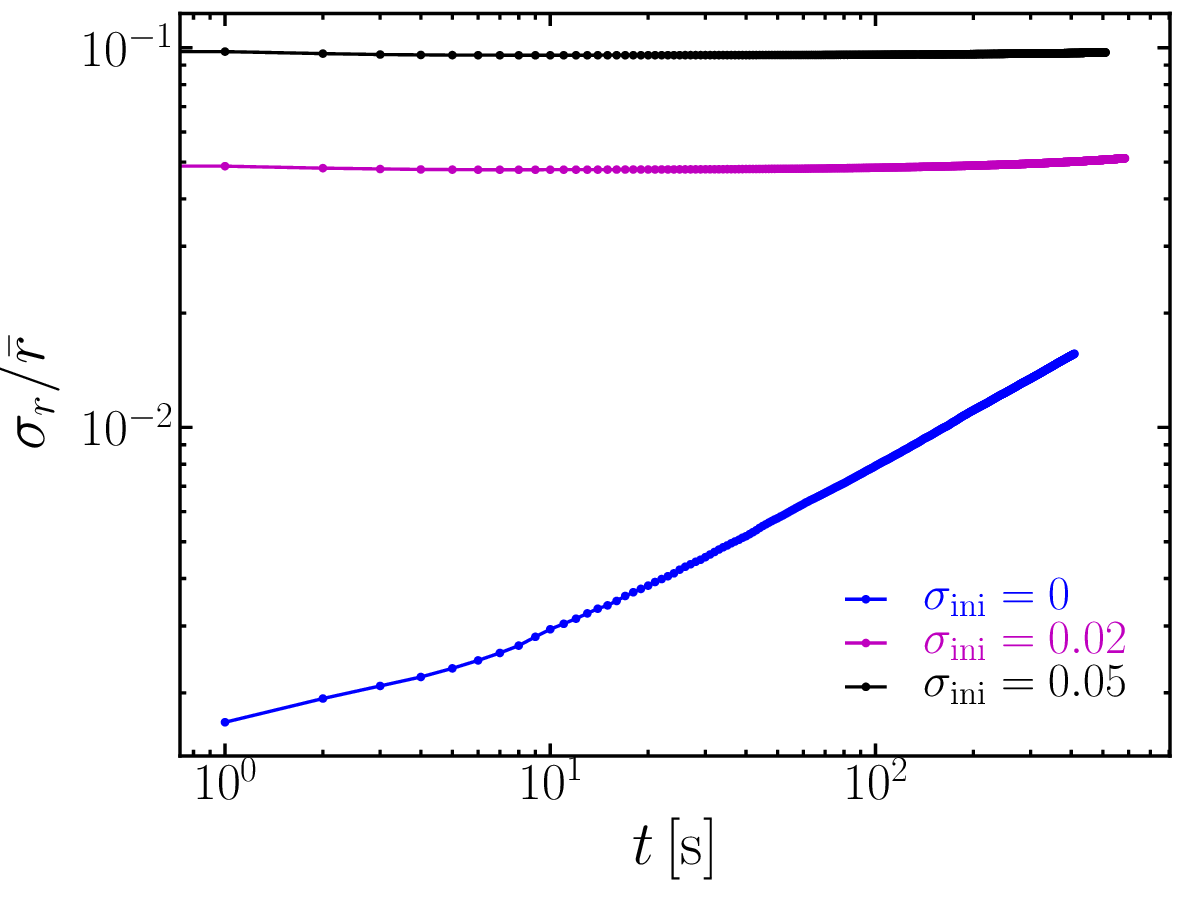}
\end{center}\caption{$\sigma_r/\bar{r}$ for different $\sigma_{\rm ini}$,
where the growth is solely driven by condensation.
}
\label{sigma_moments_cond_coll0}
\end{figure}

To investigate why $\sigma_r/\bar{r}$ is insensitive to $\sigma_{\rm ini}$,
we examine how the condensation process responds to $\sigma_{\rm ini}$.
It is evident that the condensation process is damped when
$\sigma_{\rm ini}\ge0.02$, as shown in \Fig{sigma_moments_cond_coll0}.
This suggests that condensation makes the combined processes almost independent
of the initial size distribution,
which counteracts the initial width-dependency of the collision--coalescence
process \citep{li2017effect}.

\end{document}